\documentclass[a4paper,10pt]{article}
\usepackage{graphicx}

\begin{document}

\setlength{\oddsidemargin} {1cm}
\setlength{\textwidth}{18cm}
\setlength{\textheight}{23cm}

\title{Motion equations for relativistic particles in an external electromagnetic field in scale relativity}

\author{{Marie-No\"elle C\'el\'erier and Laurent Nottale} \\
{\small Laboratoire Univers et TH\'eories (LUTH), Observatoire de Paris, CNRS \& Universit\'e Paris Diderot} \\
{\small 5 place Jules Janssen, 92190 Meudon, France} \\
{\small e-mail: marie-noelle.celerier@obspm.fr}}

\maketitle

\begin{abstract}

Klein-Gordon and Dirac equations are the motion equations for
relativistic particles with spin 0 (so-called scalar particles) and 1/2
(electron/positron) respectively. For a free particle, the Dirac equation is derived from the Klein-Gordon equation by taking its square root in a bi-quaternionic formalism fully justified by the first principles of the scale relativity theory. This is no more true when an external electro-magnetic field comes into play. If one tries to derive the electro-magnetic Dirac equation in a manner analogous to the one used when this field is absent, one obtains an additional term which is the relativistic analogue of the spin-magnetic field coupling term encountered in the Pauli equation, valid for a non-relativistic particle with spin 1/2. There is however a method to recover the standard form of the electro-magnetic Dirac equation, with no additional term, which amounts modifying the way both covariances involved here, quantum and scale, are implemented. Without going into technical details, it will be shown how these results suggest this last method is based on more profound roots of the scale relativity theory since it encompasses naturally the spin-charge coupling.

\end{abstract}

\section{Introduction}

Scale relativity allows one to give foundations to the postulates \cite{NC07} and motion equations \cite{LN93,LN96,CN03,CN04,CN06} of quantum mechanics, and to gauge theories of particle physics \cite{NCL06}, in particular to electromagnetism \cite{LN96,LN03}, by providing them with a geometric interpretation in the framework of a fractal space-time.

Successive velocity doublings proceeding from the giving up of the differentiability assumption allows one to obtain the motion equations as geodesic equations of this fractal space-time, thus implementing quantum covariance.

We have therefore been able to derive successively: the Schr\"odinger equation \cite{LN93}, which applies to non-relativistic particles, the free Klein-Gordon \cite{LN96} and Dirac \cite{CN03,CN04} equations and finally the Pauli equation \cite{CN06}, which is the non-relativistic limit of the Dirac equation and which applies to the behavior of a  non-relativistic spin 1/2 particle, e.g. a non-relativistic electron. We will come back in more details in the following to the two equations of interest here, Klein-Gordon and Dirac.

Moreover, the relativity principle applied to scales implies that the scale of an internal structure on a fractal geodesic is modified by a displacement in space-time and reciprocally. This property allows one to construct the electromagnetic field and the electric charge while giving them a physical meaning, thus implementing scale covariance.

\section{Quantum covariance and scale covariance}

Recall that in scale relativity quantum covariance is implemented by the use of a covariant velocity operator, $\widehat{{\cal V}^{\mu}}$. Owing to a first symmetry breaking, that of ${\rm d}s \leftrightarrow - \, {\rm d}s$, this operator becomes complex. When one adds the breakings of ${\rm d}x^{\mu} \leftrightarrow -\, {\rm d}x^{\mu}$ and of $x^{\mu} \leftrightarrow -\, x^{\mu}$, one gets a bi-quaternionic operator.

Being complex or bi-quaternionic, these operators are used to write the motion equations of free particles under the form of geodesic equations
\begin{equation}
\widehat{{\cal V}^{\mu}} \partial_{\mu}{\cal V}_{\nu} \left( = \frac{\widehat{\rm d}}{{\rm d}s} {\cal V}_{\nu} \right) = 0.
\end{equation}

A complex velocity operator inserted in the above equation allows one to recover the usual free Klein-Gordon equation \cite{LN96}. Recall this applies to the motion of a relativistic spinless particle non-submitted to an external field. With a bi-quaternionic velocity operator, one obtains a bi-quaternionic Klein-Gordon-like equation. The free Dirac equation follows as its square root \cite{CN03,CN04}. It applies to a relativistic spin 1/2 particle, such as the electron and its anti-particle, the positron, non-submitted to an external field. These two particles are represented, in standard quantum mechanics, by only one object, the bi-spinor, which has all the mathematical properties of a bi-quaternion. This shows that the scale relativity formalism is perfectly appropriate to reproduce phenomena observed in microphysics.

In the framework of electromagnetism, scale relativity identifies gauge transformations, whose nature was previously unknown, with global scale transformations, $\varrho = \lambda/\epsilon \rightarrow \varrho' = \lambda/\epsilon'$, in ``scale space''. It allows also to recover, while giving it a geometric interpretation as a scale covariant derivative, the usual form of the covariant derivative of quantum electrodynamics, i.e.
\begin{equation}
D_{\mu} = \partial _{\mu} + {\rm i} (e/\hbar c) A_{\mu}.
\end{equation}

Eventually, charges emerge naturally as conserved quantities in scale transformations.

Up to there, everything is perfect as regards scale relativity.

\section{Where things start to get heated}

Our aim is now to recover, from the first principles of scale relativity, both equations Klein-Gordon and Dirac, but for particles submitted to an external electromagnetic filed. The first idea which crosses our mind is to combine both tools, scale covariant derivative (QED) and quantum covariant velocity operator, by writing the strongly covariant geodesic equation:
\begin{equation}
\widehat{{\cal V}^{\mu}} D _{\mu}{\cal V}_{\nu} = 0.
\end{equation}
We obtain actually, when integrating this equation, the usual Klein-Gordon equation for a relativistic spinless particle in an external electromagnetic filed.

Proud of this success, we hoped then to be able to recover the electromagnetic Dirac equation by applying exactly the same method which was so successful in the case of free particles, i.e. extracting the square root of the bi-quaternionic electromagnetic Klein-Gordon-like equation.

More exactly, we had shown in \cite{CN03,CN04} that a bi-quaternionic free Klein-Gordon-like equation amounts to applying twice to the wave function the time part of the Dirac equation and then to equal it to its squared spatial part.

Now, this is no more the case when an electromagnetic field comes into play. An additional term appears and we have shown it corresponds to the coupling between the intrinsic magnetic moment, or spin, of the electron and the magnetic field. This term is the relativistic analogue of that implying the electron magnetic moment in the Pauli equation.

The conclusion is this method is not well-adapted and we must use another one.

\section{Correct method}

Instead of implementing a strong covariance, as in the former method, we are going to use both tools, quantum covariance and scale covariance, as in standard quantum mechanics.

First, we apply the quantum covariance and we obtain the free Klein-Gordon and Dirac equations as before. Then, we replace in these equations the ordinary derivative, $\partial_{\mu}$, by its ``inertial'' scale covariant part, $D_{\mu}$.

We obtain thus the electromagnetic Dirac equation under its usual form, where the spin-magnetic field coupling does not appear explicitly.

\section{Discussion}

We have seen that the only quantum covariant geodesic equation yields the free Dirac equation. When we add the scale covariance directly at the level of the geodesic equation, we obtain an additional term representing the spin-magnetic field coupling.

The explanation is as follows. In scale relativity, spin proceeds directly from the fractality of space-time, while charges, which stem from transformations in scale space, are only indirect consequences of this fractality. Hence, the influence of fractality through spin is more fundamental and must be applied first, independently from that of the field.

\subsection{Spin}

Spin is an intrinsic quantum property of the particles, component of the total angular momentum which is itself a constant of motion. It takes only integer or half-integer values of $\hbar$.

In scale relativity, the spin 1/2 can have two different interpretations. In the first, one considers it as a quantum charge of the electron/positron. In this case, a term representing the coupling of this charge to the external field appears explicitly in the equations. This interpretation corresponds to the way spin stems when the first strongly covariant method is used to obtain the motion equations.

\begin{figure}
\begin{center}
\includegraphics[scale=1.1]{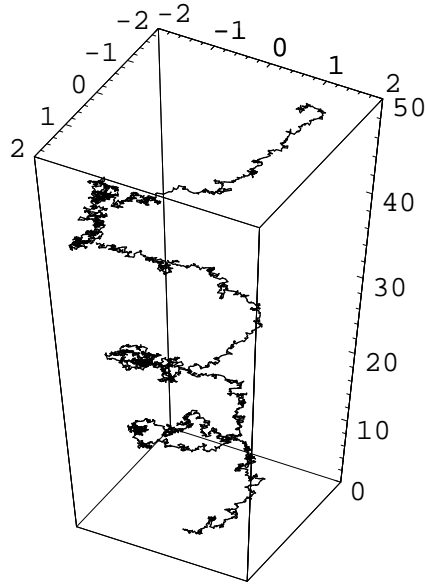}
\caption{Typical spinorial geodesic ($D_F = 2$) in a fractal space. Realization chosen between an infinity of possible realizations.}
\label{fig1}
\end{center}
\end{figure}

It can also be considered as directly linked to the fractal geometry of space-time. In this case, its coupling to the magnetic field is implicit and there is no additional term in the Dirac equation which recovers its usual form, as when the second method is used. This interpretation has been illustrated with numerical simulations aiming at visualizing the typically spinorial form of some geodesics in a fractal space-time (see Fig.~\ref{fig1}) \cite{CN06}.

\subsection{Electric charge}

Owing to the principle of relativity of scales, scale variables are explicit functions of the space-time coordinates. This implies, for every displacement in such a space-time, the appearance of a change in scale interval due to the fractal geometry, which reads
\begin{equation}
\delta \ln \varrho = \; (1/q)\;A_{\mu }\;{\rm d}x^{\mu},
\end{equation}
where $q$ is the (active) electric charge and $A_{\mu }$ defines the electromagnetic field.

It can be shown that the passive electric charge $e$ is equal to the active electric charge $q$ \cite{NCL06,LN03}.

The electric charge, a mere property of the electron stemming from scale transformations, must be considered therefore of a less fundamental nature than the spin, itself a component of a first integral of motion.

\section{Conclusion}

We have seen that the internal nature of spin, intrinsically linked to the fractal geometry of space-time is more fundamental than that of the charges, which also stem from this fractality, but less directly.

This property imposes the order of application of the two covariances associated to recover the usual electromagnetic Dirac equation:

\begin{enumerate}

\item spin

\item electric charge

\end{enumerate}

Therefore, strong covariance, which can be properly applied to recover the electromagnetic Klein-Gordon equation, since this equation does not imply spin, must be replaced by a weaker covariance in the case of the Dirac equation where spin is involved.

\end{document}